\documentstyle[12pt,axodraw]{article}
\begin{document}
\newcommand{\bfig}{\begin{center}\begin{picture}}
\newcommand{\efig}[1]{\end{picture}\\{\small #1}\end{center}}
\newcommand{\flin}[2]{\ArrowLine(#1)(#2)}
\newcommand{\wlin}[2]{\DashLine(#1)(#2){2.5}}
\newcommand{\zlin}[2]{\DashLine(#1)(#2){5}}
\newcommand{\glin}[3]{\Photon(#1)(#2){2}{#3}}
\newcommand{\lin}[2]{\Line(#1)(#2)}
\newcommand{\sof}{\SetOffset}
\newcommand{\bmip}[2]{\begin{minipage}[t]{#1pt}\bfig(#1,#2)}
\newcommand{\emip}[1]{\efig{#1}\end{minipage}}
\newcommand{\putk}[2]{\Text(#1)[r]{$p_{#2}$}}
\newcommand{\putp}[2]{\Text(#1)[l]{$p_{#2}$}}
\newcommand{\bq}{\begin{equation}}
\newcommand{\eq}{\end{equation}}
\newcommand{\bqa}{\begin{eqnarray}}
\newcommand{\eqa}{\end{eqnarray}}
\newcommand{\nl}{\nonumber \\}
\newcommand{\eqn}[1]{eq. (\ref{#1})}
\newcommand{\eqs}[1]{eqs. (\ref{#1})}
\newcommand{\ibidem}{{\it ibidem\/},}
\newcommand{\vpb}{}

\title{
\vspace{-4cm}
\begin{flushright}
{\large  PSI-PR-96-19} \\
\end{flushright}
\vspace{4cm}
A simple method for multi-leg loop calculations}
\author{R.~Pittau\thanks{email address: pittau@psw218.psi.ch}\\
        Paul Scherrer Institute, CH-5232 Villigen-PSI, Switzerland }
\maketitle
\thispagestyle{empty}
\begin{abstract}
In this paper, I present a technique to simplify the tensorial reduction
of one-loop integrals with arbitrary internal masses, but at least 
two massless external legs. 
By applying the method to rank $l$ tensor integrals, one ends up with
at most rank 1 tensor functions with the initial number
of denominators, plus tensor integrals with less denominators and rank $< l$.
To illustrate the algorithm, I explicitly compute diagrams contributing to
processes of physical interest and show how the usual numerical
instabilities due to the appearance of Gram determinants can be controlled.
\end{abstract}
\clearpage
\section{Introduction}
Present and future experiments in high energy physics require an
increasing understanding of multi-particle final states.
Idealized calculations, in which heavy particles are treated as being
on-shell, are often not accurate enough to match the experimental needs.
In fact, the produced heavy particles always decay, giving a multi-body
final state as an observable signal. 

\noindent At first, investigations of multi-particle processes can be
performed at the tree level, but, in a second stage, also radiative
corrections have to be taken into account. At LEP2, for instance, a complete
electroweak one-loop calculation to off-shell four-fermion production 
beyond the leading log approximation is still missing, and this reflects on
the accuracy in the $W$ mass measurement. Besides, next-to-leading
order QCD calculations of multi-jet final states are becoming unavoidable 
to control the background in the search of new signals. 
Therefore, it is clear that multi-leg loop calculations
are playing a role of increasing importance and that any progress
achieved in this field is welcome.\\

One of the major problems in loop calculations is the 
reduction (in $n$ dimensions) of rank $l$ tensor $m$-points integrals 
as follows
\bqa
I_m^{\mu_1 \cdots \mu_l}&=& \int d^nq~\frac {q^{\mu_1}\cdots q^{\mu_l}}
{D_1 \cdots D_m} \nl
D_i &=& (q+s_{i-1})^2 -m^2_i \nl
s_i &=& \sum_{j=\,0}^{i} k_j ~~~~~~k_0= 0\,.
\label{eq0}
\eqa
The standard reduction method \cite{pasve} 
consists in decomposing the tensorial 
structure in terms of momenta $k_j$ and metric tensor. For
example, a rank two 5-point one-loop integral can be expressed in
terms of $11$ linearly independent scalar form factors
\bqa
I_5^{\mu\nu}&=& \int d^nq~\frac {q^{\mu}\,q^{\nu}}{D_1\, D_2\, 
           D_3\, D_4\, D_5} \nonumber \\
&=& E_{21}~k_1^{\mu}\,k_1^{\nu}+E_{22}~k_2^{\mu}\,k_2^{\nu}+ \cdots +
  E_{211}~g^{\mu\nu} \,. 
\label{eq1}                           
\eqa
Analogously, the rank three 5-point function $I_5^{\mu\nu\rho}$ 
gives $24$ terms.

\noindent The freedom in choosing the basis for the tensorial
decomposition can be used to define suitable 
combinations $v_j$ of $k_j$ such that $v_i \cdot k_j= \delta_{ij}$ 
\cite{vanold}. 
In any case, in order to express the coefficients of
the decomposition in terms of scalar loop functions, one has to solve
algebraical systems obtained by multiplying both sides of \eqn{eq0}
by $k_j$ (or linear combinations of them) and the metric tensor.
This causes the appearance of Gram determinants 
\bqa
 \Delta= {\rm det}(k_i \cdot k_j) \nonumber
\eqa
into the denominator, that can give rise to numerical instabilities.
In fact, they vanish in collinear regions of the phase space where 
the cross section is well defined, so that large numerical
cancellations are expected in the numerator 
in order to keep everything finite.

\noindent Different approaches also exist in which the form
factors are identified with integrals in a different number of
dimensions \cite{davy} or directly formulated in the Feynman parameter
space \cite{bern0}.
In both cases, Gram determinants still appear at some stage in the
reduction.

\noindent Recently, a new method has been proposed by Campbell, Glover 
and Miller \cite{glover}, in which the coefficients of the reduction are 
built up by combining the scalar integrals in
groupings that are well behaved in the limit $\Delta \to 0$. This
somehow solves the problem of the Gram determinants, but to actually
construct the well behaved groupings can be cumbersome in general 
- non QCD like - cases, especially when the rank of the appearing tensor
integrals is large. Furthermore, the number of the basic functions
increases, and additional scalar integrals must be evaluated
in $n=6+\epsilon$ or higher dimensions.\\ 

In this paper, I propose a method to simplify Feynman diagrams, 
in which no evaluation of new functions is required, apart from the
usual scalar integrals (new functions may appear when $n= 4+\epsilon$,
but their computation is trivial). Although, in general, the appearance 
in the denominator of quantities that can vanish in some corner of the
phase space cannot be completely avoided, a better control 
of those singularities is possible with respect to the traditional
methods of refs. \cite{pasve,vanold}. 
The procedure is somehow complementary to the technique of
Campbell, Glover and Miller, and can be used to simplify the problem
before applying their method.\\

A trivial example serves to illustrate the basic idea. 
Consider the tensorial reduction of the following quantity
\bqa
\gamma_\mu\,\gamma_\nu I_5^{\mu\nu} = 
\int d^nq~\frac{\rlap{/} q \,\rlap{/} q}{D_1\, D_2\, 
           D_3\, D_4\, D_5}~.
\label{eq2} 
\eqa
One can certainly use \eqn{eq1},
but the equation 
\bqa
\rlap{/} q \, \rlap{/} q= q^2= D_1+m_1^2 \nonumber
\label{eq3}
\eqa
immediately gives the desired answer in terms of just two scalar functions
\bqa
\gamma_\mu\,\gamma_\nu I^{\mu\nu} = \int d^nq~\frac{1}{D_2\, 
           D_3\, D_4\, D_5} 
                           + m_1^2\,\int d^nq~\frac{1}{D_1\, D_2\, 
           D_3\, D_4\, D_5}\,. 
\eqa
Therefore, by using in a diagram the algebra of the $\gamma$ matrices  
to reconstruct denominators, one gets a simpler tensorial structure.

\noindent This technique of reconstructing denominators has been already
used in the literature for the evaluation of specific integrals 
\cite{bern,kk}. 
In this paper, I show that a reduction based on this procedure can
always be worked out for 
generic one-loop integrals with arbitrary internal masses, but at least 
two massless external legs.\\

In the next section, I present the method, with the help two 
4-dimensional examples. In section 3 and appendix A, 
I describe the extension to 
$n= 4+\epsilon$ dimensions and give explicit results for cases of 
physical interest. 

\section{The Method}
After the decay of the intermediate heavy particles,
a generic $m$-point loop diagram can be written as follows (see fig. 1-a)
\bqa
{\cal M}(p_1,\cdots,p_r;k_1, \cdots , k_{m-1}) =
\sum_i^{N}\int\,d^n q 
\frac{Tr^{(i)}\,[\rlap{/}q \cdots 
\rlap{/}q \cdots\,]}{D_1\,\cdots D_m} \equiv \sum_i^{N} {\cal M}^{(i)}\,,
\label{eqgen}
\eqa 
where $p_{1 \cdots\, r}$ are the $r$ external momenta {\em of the diagram}, 
$k_{1 \cdots\, m-1}$ the momenta {\em in the loop functions} - as 
defined in \eqn{eq0} - and $Tr^{(i)}$ traces
over $\gamma$ matrices, which contain $\rlap{/} q$'s.
In most practical cases, the $r$ external momenta are massless, 
in fact they can only be photons, gluons or light fermions.
The appearance of traces in \eqn{eqgen} is a general feature. In fact,
if the external particles of the diagram are fermions, traces naturally
appear, while the polarization vectors of photons and gluons can always
be written in terms of spinor strings using the following 
representation \cite{kleiss}
\bqa
\epsilon^\mu(\lambda,k)= \frac{1}{\sqrt 2} \frac{\bar v_{-\lambda}(k)
\gamma^\mu u_{\lambda}(b)}{\bar v_{\lambda}(b)
u_{\lambda}(k)}~~~~~~~ b\,=\,{\rm  arbitrary~massless~vector\,.}
\label{eqvec}
\eqa
\noindent In standard calculations, each term in \eqn{eqgen}
is rewritten as follows
\bqa
{\cal M}^{(i)}(p_1,\cdots,p_r;k_1, \cdots , k_{m-1}) = \int\,d^n q 
\frac{q^\alpha \cdots q^\beta \cdots \,Tr^{(i)}\,[\gamma_\alpha \cdots 
\gamma_\beta \cdots\,]}{D_1\,\cdots D_m}
\label{eqgen1}
\eqa
and the tensors
\bqa
\int\,d^n q \frac{q^\alpha \cdots q^\beta \cdots}{D_1\,\cdots D_m}
\label{eqtens}
\eqa
decomposed using a basis of independent four-vectors, as described in
the introduction. Instead, I propose to use $\gamma$-algebra and spinor
manipulations to move all $\rlap{/}q$'s close to each other, and subsequently 
apply $\rlap{/}q\rlap{/}q= q^2$ in order to rewrite each trace as
\bqa
Tr^{(i)}\,[\rlap{/}q \cdots 
\rlap{/}q \cdots\,]= \sum_k A_k\, Tr^{(i)}_k [\rlap{/}q \cdots]
+ \sum_l B_l\, Tr^{(i)}_l [\cdots]\,,
\label{eqgen2}
\eqa
where $Tr^{(i)}_k$ contain at most one $\rlap{/}q$, $Tr^{(i)}_l$
do not contain $\rlap{/}q$ and the coefficients $A_k$ and $B_l$ are
functions of the scalar products between $q$ and the external momenta
{\em of the diagram } $p_{1 \cdots r}$.
\noindent Since a single $\rlap{/} q$ appears in the right 
side of \eqn{eqgen2}, 
just a rank 1 tensor decomposition is formally left. Anyway,
there is no gain in applying the described procedure if the coefficients 
$A_k$ and $B_l$ are generic. In fact, if scalar products of the kind 
$(q \cdot p_j)$ appear and $p_j$ is not equal to one of the momenta 
$k_j$ {\em in the loop functions},
the same kind of tensorial decomposition necessary for computing 
\eqn{eqtens} is hidden in \eqn{eqgen2}. 
On the other hand, if only powers of $(q \cdot k_j)$ and $q^2$ appear, it is
always possible to express at least one of them in terms of denominators.
In this case, starting from $m$-point rank $l$ tensor integrals, 
the algorithm gives at most rank 1 $m$-point functions,
plus $n$-point rank $p$ tensor integrals with $n < m$ and $p < l$.

\noindent Naturally the question arises whether 
the coefficients $A_k$ and $B_l$ can
be made to have such a nice feature. To answer that question, while keeping 
everything transparent, I will assume, for the rest of this section, 
four-dimensional space time, namely ultraviolet finite integrals, 
with four-dimensional regularization of infrared and collinear
divergences. The extension to $n= 4+\epsilon$ dimensions will 
be considered in the next section. In the following, 
I introduce all notations and necessary formulae.\\

\noindent 
Basic objects are strings of $\gamma$ matrices
between massless spinors. I will use the notation
\bqa
\bar v_+(i)\, \rlap{/} p_3 \,\rlap{/} p_4 \,\rlap{/} q \,\rlap{/} p_5
  \cdots \, \rlap{/} p_6\, u_-(j)
 = \left\{3\,4\,q\,5\,\cdots\,6\right\}^{+-}_{i~j}\,.
\label{eq4}
\eqa 
The Weyl spinors are defined as 
\bqa
  u_\pm(j)= \Pi_\pm \,u(j)~~~&&~~~~\bar u_\pm(j)= \bar u\, \Pi_\mp \nl 
  v_\pm(j)= \Pi_\mp \,v(j)~~~&&~~~~\bar v_\pm(j)= \bar v\, \Pi_\pm \,,  
\label{eq5}
\eqa
where $j$ denotes the spinor momentum and 
$\Pi_\pm = \frac{1}{2}(1 \pm \gamma_5)$ are chirality projectors
obeying
\bqa
\Pi_+ \, \Pi_- =  0~~~~(\Pi_\pm)^2 = \Pi_\pm\,.
\label{eq6}
\eqa
Note that, using $u_\pm(j)= v_\mp(j)$, one can always 
rewrite any spinorial string with $v$  
spinors on the left and $u$ spinors on the right, so that the notation
in \eqn{eq4} is completely general.

\noindent The completeness relations for massless spinors read
\bqa
u_{\pm}(j) \bar u_{\pm}(j)&=& \rlap{/}  p_j\,  \Pi_{\mp} \nl
v_{\pm}(j) \bar v_{\pm}(j)&=& \rlap{/}  p_j\,  \Pi_{\pm}\,,
\label{eq8}
\eqa
from which one can easily get similar relations for
generic momenta $i$ and $j$ \cite{passa}
\bqa
u_-(i) \bar v_-(j)&=& -\frac{\rlap{/} p_i \rlap{/} p_j} 
{<i\,j>}\, \Pi_- \nl
u_+(i) \bar v_+(j)&=&  \frac{\rlap{/} p_i \rlap{/} p_j} 
{<i\,j>^\ast}\, \Pi_+ \nl
u_-(i) \bar v_+(j)&=&  \frac{\rlap{/} p_i \rlap{/} b \rlap{/} p_j}
{\bar v_+(i)\, \rlap{/} b u_-(j)} \, \Pi_+\nl 
u_+(i) \bar v_-(j)&=&  \frac{\rlap{/} p_i \rlap{/} b \rlap{/} p_j}
{\bar v_-(i)\, \rlap{/} b u_+(j)} \, \Pi_-\,, 
\label{eq9}
\eqa
where $b$ is an arbitrary four-vector and 
\bqa
\begin{array}{lcrcr}
<i j> ^\ast &=&   \bar v_-(i)\,u_-(j)&=&  \bar u_+(i)\,v_+(j) \\\\
<i j>       &=&  -\bar v_+(i)\,u_+(j)&=&- \bar u_-(i)\,v_-(j) 
\label{eq10}
\end{array}
\eqa
are scalar products in the spinor space \cite{kleiss,hely}.

\noindent The following identities hold for spinor strings \cite{korner1} 
\bqa
\left\{\Gamma   \right\}^{+-}_{i~j} &=& 
\left\{\tilde \Gamma\right\}^{-+}_{j~i}     \nl
\left\{\Gamma   \right\}^{\pm \pm}_{i~j}&=& -
\left\{\tilde \Gamma\right\}^{\pm \pm}_{j~i}\,,
\label{eq7}
\eqa
$\tilde \Gamma$ being the string obtained from $\Gamma$ by reversing
the order of the $\gamma$ matrices.

\noindent The following Kahane-Chisholm identities \cite{kahane}
are also very useful
\bqa
\gamma_\alpha \Gamma_o \gamma^\alpha = -2 \,\tilde{\Gamma}_o~~~~~~~~~~
\gamma_\alpha \Gamma_e \gamma^\alpha = 
      Tr[\Gamma_e]-\gamma_5\, Tr[\gamma_5\Gamma_e]\,,
\label{eq11}
\eqa
where $\Gamma_o$ and $\Gamma_e$ stand for strings
with an odd and even number of $\gamma$ matrices, respectively.
Finally, one can prove \cite{korner1}
\bqa
\rlap{/} b \Gamma_e \rlap{/} b= - \rlap{/} b \tilde{\Gamma}_e \rlap{/} b
{\rm~~~~~~~~~~if~~} b^2= 0 \,.
\eqa
 
\vspace {0.3cm}

\noindent Armed with the previous formulae, I will show that,
in four dimensions, the coefficients $A_k$ and $B_l$ in \eqn{eqgen2} can
be made to depend only on powers of $q^2$ and $(q \cdot k_j)$
{\em on the condition that at least two momenta in the
set} $k_{1\cdots m-1}$ (say $k_1$ and $k_2$) {\em are massless}, namely 
coincide with some of the external massless momenta of the diagram, 
as in fig. 1-b and 1-c.

\noindent In fact, the following identity 
\bqa
 \rlap{/} q = \frac{1}{2\,(k_1 \cdot k_2)}\,\left[
               2\,(q \cdot k_2)\,\rlap{/} k_1   
              +2\,(q \cdot k_1)\,\rlap{/} k_2 
              -\rlap{/} k_1 \rlap{/} q \rlap{/} k_2
              -\rlap{/} k_2 \rlap{/} q \rlap{/} k_1 \right]
\label{eqext}
\eqa
\bfig(300,140)
\SetScale{1}
\sof(0,60)
\flin{-10,50}{10,10} 
\flin{10,10}{30,50} 
\Gluon(10,10)(55,25){2}{6}
\Gluon(10,10)(55,-5){2}{6}
\glin{10,10}{30,-30}{10}
\flin{-20,-30}{10,10} 
\Text(-10,56)[rb]{$p_1$}
\Text(30,56)[lb]{$p_2$}
\Text(61,25)[l]{$p_3$}
\Text(61,-5)[l]{$p_4$}
\Text(31,-33)[tl]{$p_5$}
\Text(-20,-33)[tr]{$p_r$}
\Text(15,-21)[tr]{{\Large $\cdots$}}
\GOval(10,10)(10,10)(0){0.9}
\Text(15,-50)[t]{{\em a}}
\sof(120,60)
\flin{-10,50}{10,10} 
\flin{10,10}{30,50} 
\glin{-5,40}{25,40}{6}
\Gluon(10,10)(55,25){2}{6}
\Gluon(10,10)(55,-5){2}{6}
\glin{10,10}{30,-30}{10}
\flin{-20,-30}{10,10} 
\Text(-10,56)[rb]{$p_1$}
\Text(30,56)[lb]{$p_2$}
\Text(61,25)[l]{$p_3$}
\Text(61,-5)[l]{$p_4$}
\Text(31,-33)[tl]{$p_5$}
\Text(-20,-33)[tr]{$p_r$}
\Text(15,-21)[tr]{{\Large $\cdots$}}
\GOval(10,10)(10,10)(0){0.9}
\Text(15,-50)[t]{{\em b}}
\sof(240,60)
\flin{-10,50}{10,10} 
\flin{10,10}{30,50} 
\Gluon(10,10)(55,25){2}{6}
\Gluon(10,10)(55,-5){2}{6}
\Gluon(43,0)(43,19){2}{3}
\glin{10,10}{30,-30}{10}
\flin{-20,-30}{10,10} 
\Text(-10,56)[rb]{$p_1$}
\Text(30,56)[lb]{$p_2$}
\Text(61,25)[l]{$p_3$}
\Text(61,-5)[l]{$p_4$}
\Text(31,-33)[tl]{$p_5$}
\Text(-20,-33)[tr]{$p_r$}
\Text(15,-21)[tr]{{\Large $\cdots$}}
\GOval(10,10)(10,10)(0){0.9}
\Text(15,-50)[t]{{\em c}}
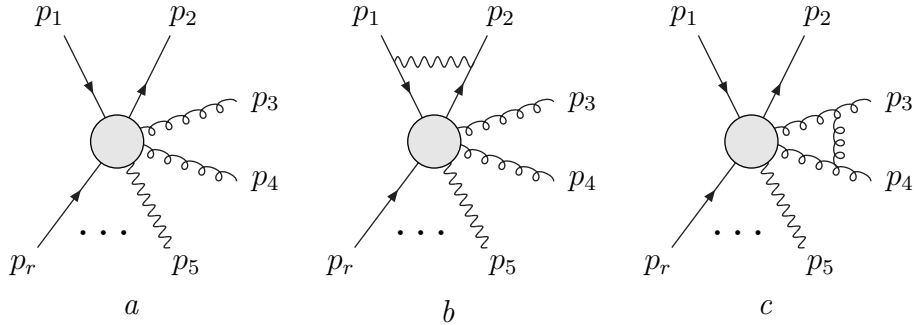
\efig{Figure 1: {\em Generic one-loop multi-leg process (a). 
The blob stands for the sum of all possible one-loop diagrams. 
Two classes of diagrams are also shown, for which the 
reduction method is applicable (b and c).\\}}
allows to ``extract'' $q$ from the traces 
\bqa
Tr[\Gamma_1 \rlap{/} q \Gamma_2] &=& \frac{1}{2\,(k_1 \cdot k_2)}
      \{ 
          2\,(q \cdot k_2)\,Tr[\Gamma_1 \rlap{/} k_1 \Gamma_2]
         +2\,(q \cdot k_1)\,Tr[\Gamma_1 \rlap{/} k_2 \Gamma_2] \nl
         &-&Tr[\Gamma_1 \rlap{/} k_1 \rlap{/} q \rlap{/} k_2 \Gamma_2]
         -Tr[\Gamma_1 \rlap{/} k_2 \rlap{/} q \rlap{/} k_1 \Gamma_2]
      \}\,,
\label{eqden}
\eqa
where $\Gamma_{1,2}$ represent generic strings of $\gamma$
matrices. In the first two terms, the $q$ dependence is already
extracted out. The other two traces can be broken with the help of \eqn{eq8}
\bqa
Tr[\Gamma_1 \rlap{/} k_1 \rlap{/} q \rlap{/} k_2 \Gamma_2] &=&
Tr[\Gamma_1 \rlap{/} k_1 \rlap{/} q \rlap{/} k_2(\Pi_+ + \Pi_-) \Gamma_2] \nl
&=& \left\{q \right\}^{+-}_{1~2} \left\{\Gamma_2 \Gamma_1 \right\}^{+-}_{2~1}
   +\left\{q \right\}^{-+}_{1~2} 
    \left\{\Gamma_2 \Gamma_1 \right\}^{-+}_{2~1}\,.
\eqa
By iteratively applying the above procedure together with 
the identities (derived again by means of \eqn{eq8} and \eqn{eq9})
\bqa
\left\{q \right\}^{-+}_{1~2} \left\{q \right\}^{-+}_{2~1}
  &=& 4\,(q \cdot k_1)\,(q \cdot k_2) -2\,q^2\,(k_1 \cdot k_2) \nl
\left\{q \right\}^{-+}_{1~2} \left\{q \right\}^{-+}_{1~2}
  &=& \frac{1} {\bar v_+(1)\, \rlap{/} b u_-(2)}
\left[  2\,(q \cdot k_1)\left\{b2q \right\}^{+-}_{2~1}
       -2\,(q \cdot b)\left\{12q \right\}^{+-}_{2~1} \right. \nl
  &+&\left. 2\,(q \cdot k_2)\left\{1bq \right\}^{+-}_{2~1}
            -q^2\left\{1b2 \right\}^{+-}_{2~1} \right]
\label{eqext1}
\eqa
it is easy to see that the resulting 
coefficients $A_k$ and $B_l$ only depend on
powers of $q^2$, $(q \cdot k_1)$, $(q \cdot k_2)$ and 
$(q \cdot b)$.
The vector $b$ is arbitrary (and not necessarily massless), but
different from $k_1$ and $k_2$. By identifying it with a third
momentum in the set $k_{1 \cdots m-1}$
\footnote{This identification is possible only for $m$-point loop
functions with $m > 3$. However, in the next section, I will show that
the method still works when $m\le 3$.},
$A_k$ and $B_l$ have the suitable structure to get
the desired simplifications in the tensorial decomposition.\\

In the computation of physical processes, 
the one-loop diagrams with highest $m$ always lie in the class of 
the corrections connecting two external legs (see again fig. 1-b and 1-c). 
That guarantees at least two massless momenta in the loop functions
and, therefore, that 
the described reduction method is applicable at least for the most
complicated diagrams appearing in the calculation.

\noindent Finally, one should notice that structures that can vanish appear
in the denominator of eqs. (\ref{eqden}) and (\ref{eqext1}). 
However, as we will see, often in practical calculations there is no
need to extract the $q$ dependence as described, since the
reconstruction of denominators naturally takes place without 
introducing too many factors in the denominators.

\noindent In the following two examples, I show the method at work in 
four dimensions.
\vspace{0.3cm}

\noindent{\large\em Example 1}

\vspace{0.3cm}

\noindent I shall compute, in the renormalizable gauge, the 6-point
diagram of fig. 2, relevant for studying electroweak corrections
at LEP2.

\noindent The spinorial structure in the numerator of the diagram reads
\bqa
{\cal A} &=& -\bar v_+(3)\, \gamma_\mu   \rlap{/} p_B  \gamma_\beta\,u_-(4)
        \,~   \bar v_+(5)\, \gamma^\beta \rlap{/} p_A  \gamma^\nu  \,u_-(6)
        \,~   \bar v_+(1)\, \gamma^\mu   \rlap{/} p_C  \gamma_\nu\,u_-(2) \nl
         &\equiv& - \left\{\mu B \beta \right\}^{+-}_{3~4}
         \left\{\beta A \nu \right\}^{+-}_{5~6}
         \left\{\mu C \nu   \right\}^{+-}_{1~2} \,,
\eqa
from which the scattering amplitude ${\cal M}$ can be obtained 
after division by the denominators and integration over $d^4q$ 
\bqa
{\cal M} &=& \int d^4q\, \frac{{\cal A}}{D_1\, D_2\, D_3\, D_4\,
                                 D_5\, D_6} \,. \nonumber 
\eqa
\bfig(300,130)
\SetScale{1}
\sof(-60,20)
\flin{40,0}{65,25} \flin{65,25}{65,70} \flin{65,70}{40,95}
\wlin{65,25}{110,25}
\wlin{65,70}{110,70}
\flin{145,65}{110,70}\flin{110,70}{135,95}
\flin{135,0}{110,25}\flin{110,25}{145,30}
\glin{135,29}{135,66}{8}
\Text(37,0)[tr]{$2$}
\Text(37,95)[br]{$1$}
\Text(138,95)[bl]{$3$}
\Text(149,65)[l]{$4$}
\Text(149,30)[l]{$5$}
\Text(138,0)[tl]{$6$}
\Text(126,63)[t]{\footnotesize{B}}
\Text(126,32)[b]{\footnotesize{A}}
\Text(139,47)[l]{\footnotesize{V}}
\Text(60,47)[r]{\footnotesize{C}}
\Text(87,29)[b]{\footnotesize{W}}
\Text(87,74)[b]{\footnotesize{W}}
\Text(185,95)[l]{$p_A= q~~~~~~~~~p_B= q+p_4+p_5$}
\Text(185,80)[l]{$p_C= q-p_2-p_6$}
\Text(185,65)[l]{$D_1=  q^2$}
\Text(185,50)[l]{$D_2= (q+p_5)^2-M_V^2$}
\Text(185,35)[l]{$D_3= (q+p_5+p_4)^2$}
\Text(185,20)[l]{$D_4= (q+p_5+p_4+p_3)^2-M_W^2$}
\Text(185,5)[l]{$D_5=  (q+p_5+p_4+p_3+p_1)^2$}
\Text(185,-10)[l]{$D_6=(q+p_5+p_4+p_3+p_1+p_2)^2-M_W^2$}
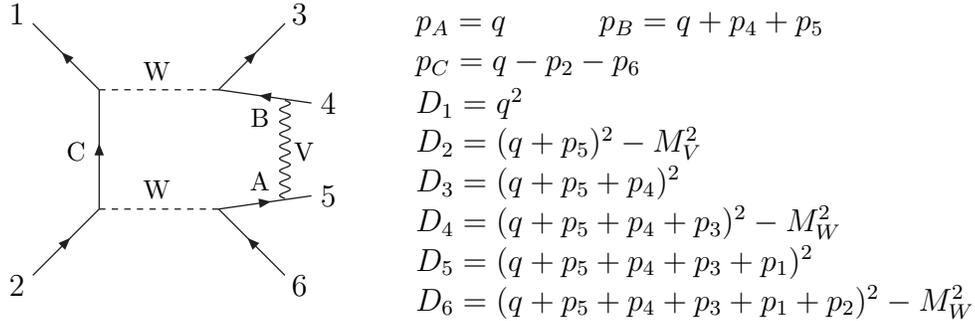
\efig{Figure 2: {\em LEP2 6-point diagram. All momenta are incoming.
V can be either a massive gauge boson or a photon. In the latter case, an
appropriate four-dimensional infrared regularization is understood.}}
Using \eqn{eq9} one can rewrite
\bqa
{\cal A} &=& -\frac{1}{N} \left\{\mu B \beta 4 b 5 \beta A \nu 6 b 1 
                                 \mu C \nu \right\}^{+-}_{3~2} \nl 
         &=& -\frac{4}{N} \left\{1 b 6 \nu A 4 b 5 
                                 B C \nu \right\}^{+-}_{3~2}   \nl
   N &=& \bar v_+(4)\, \rlap{/} b u_-(5)\,~\bar v_+(6)\,\rlap{/} b u_-(1)\,, 
\eqa
where I have applied twice the Chisholm identity in \eqn{eq11} to
sum over $\beta$ and $\mu$.
By introducing a new summed four-dimensional index $\alpha$,
the products $u_-(i) \bar v_+(j)$ can be reconstructed
\bqa
{\cal A} &=& 2\,\,\bar v_+ (3) \gamma_\alpha u_-(6)
\,\,\bar v_+(1) \gamma^\alpha
 \gamma^\nu \rlap{/} p_Au_-(4)\,\,\bar v_+(5) \rlap{/}p_B\rlap{/} p_C
 \gamma_\nu u_-(2)\,
\eqa
and, after Fierz reordering the last two strings by means of the second of
\eqs{eq11}, one gets
\bqa
{\cal A} &=& 4\,\,\bar v_+(3) \gamma_\alpha u_-(6)\,\,\bar v_+(1) 
\gamma^\alpha
 u_-(2)\,\,\bar v_+(5) \rlap{/}p_B \rlap{/}p_C \rlap{/}p_A u_-(4)\,.
\eqa
Since
\bqa
\rlap{/}p_B \rlap{/}p_C \rlap{/} p_A &=& [ D_5+
    \rlap{/}p_{13} \rlap{/}p_{26}] \rlap{/} q
    -D_1 \,\rlap{/}p_{13}  \nl
\rlap{/}p_{ij} &=& \rlap{/}p_i+\rlap{/}p_j
\eqa
denominators naturally appear in ${\cal A}$, so that the scattering amplitude 
is immediately expressed in terms of one scalar 5-point function,
one rank one 5-point function and one 6-point function or rank 1.
\vspace{0.3cm}

\noindent{\large\em Example 2}

\vspace{0.3cm}

\noindent A different contribution comes from the ``crossed''
6-point diagram in fig. 3.

\vspace{0,4cm}

\noindent
\bfig(300,130)
\SetScale{1}
\sof(-60,20)
\flin{40,0}{65,25} \flin{65,25}{65,70} \flin{65,70}{40,95}
\wlin{65,25}{110,25}
\wlin{65,70}{110,70}
\flin{110,70}{145,65}\flin{135,95}{110,70}
\flin{110,25}{135,0}\flin{145,30}{110,25}
\glin{135,29}{135,66}{8}
\Text(37,0)[tr]{$2$}
\Text(37,95)[br]{$1$}
\Text(138,95)[bl]{$4$}
\Text(149,65)[l]{$3$}
\Text(149,30)[l]{$6$}
\Text(138,0)[tl]{$5$}
\Text(126,63)[t]{\footnotesize{A}}
\Text(126,32)[b]{\footnotesize{B}}
\Text(139,47)[l]{\footnotesize{V}}
\Text(60,47)[r]{\footnotesize{C}}
\Text(87,29)[b]{\footnotesize{W}}
\Text(87,74)[b]{\footnotesize{W}}
\Text(185,95)[l]{$p_A= q~~~~~~~~~p_B= q+p_3+p_6$}
\Text(185,80)[l]{$p_C= q-p_1-p_4$}
\Text(185,65)[l]{$D_1=  q^2$}
\Text(185,50)[l]{$D_2= (q+p_3)^2-M_V^2$}
\Text(185,35)[l]{$D_3= (q+p_3+p_6)^2$}
\Text(185,20)[l]{$D_4= (q+p_3+p_6+p_5)^2-M_W^2$}
\Text(185,5)[l]{$D_5=  (q+p_3+p_6+p_5+p_2)^2$}
\Text(185,-10)[l]{$D_6=(q+p_3+p_6+p_5+p_2+p_1)^2-M_W^2$}
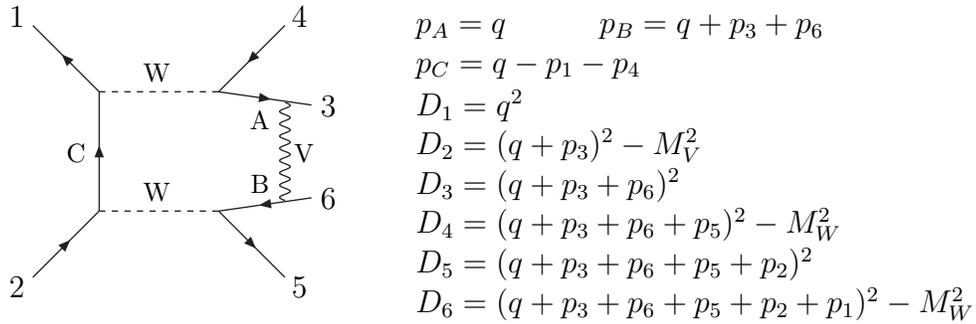
\efig{Figure 3: {\em ``Crossed'' LEP2 6-point diagram.}}
The numerator of the amplitude reads
\bqa
{\cal A} &=&   \left\{\mu C \nu \right\}^{+-}_{1~2}
        \left\{\nu B \beta \right\}^{+-}_{5~6}
        \left\{\beta A \mu   \right\}^{+-}_{3~4}\,. 
\eqa
Exactly as before, one easily performs the sums
over $\mu$, $\nu$ and $\beta$
\bqa 
{\cal A} \,\,=\,\, -8\,
        \left\{A \right\}^{+-}_{1~6}
        \left\{B \right\}^{+-}_{3~2}
        \left\{C \right\}^{+-}_{5~4} 
           =  8\, \left[\,  
        \left\{q \right\}^{+-}_{1~6}
        \left\{6 \right\}^{+-}_{3~2}
        \left\{1 \right\}^{+-}_{5~4}\right. \nl
  - \left. \left\{q \right\}^{+-}_{1~6}
            \left\{6 \right\}^{+-}_{3~2}
            \left\{q \right\}^{+-}_{5~4} 
   +        \left\{q \right\}^{+-}_{1~6}
            \left\{q \right\}^{+-}_{3~2}
            \left\{1 \right\}^{+-}_{5~4} 
   -        \left\{q \right\}^{+-}_{1~6}
            \left\{q \right\}^{+-}_{3~2}
            \left\{q \right\}^{+-}_{5~4}\,\right].
\label{eqex2}
\eqa
The first term in \eqn{eqex2} only contains one $q$ and cannot
be further manipulated, while the second and the third term can be reduced 
with the help of \eqn{eq9}. For example
\bqa 
  &&\left\{q \right\}^{+-}_{1~6} \left\{q \right\}^{+-}_{5~4} =
    \left\{q \right\}^{+-}_{1~6} \left\{q \right\}^{-+}_{4~5} =
 -\frac{1}{<64>}\,\left\{q64q \right\}^{++}_{1~5} \nl
 &=& -\frac{1}{<64>}\left[\, 
         2\,(q \cdot p_6) \left\{4q \right\}^{++}_{1~5}
        -2\,(q \cdot p_4) \left\{6q \right\}^{++}_{1~5}
        +q^2            \left\{64 \right\}^{++}_{1~5} \,\right].
\label{eqex2a}
\eqa
Similarly, one gets for the last term
\bqa
&&~~~~~~\left\{q \right\}^{+-}_{1~6} \left\{q \right\}^{+-}_{3~2} 
\left\{q \right\}^{+-}_{5~4} \,\,=\,\,
\left\{q \right\}^{+-}_{1~6} \left\{q \right\}^{-+}_{2~3}
\left\{q \right\}^{+-}_{5~4} \nl
&&~~~~~~~~~~~~~~~~~~~~~~~~~~~~~~~~~=\,\, -\,\,\frac{1}{<62><35>^\ast}\,
            \left\{q62q35q \right\}^{+-}_{1~4} \,\,=\nl 
&-&\!\!\!\frac{1}{<62><35>^\ast}\,\left[\,4\,(q\cdot p_6)\,(q\cdot p_3)\,
 \left\{25q\right\}^{+-}_{1~4} - 4\,(q\cdot p_6)\,(q\cdot p_5)\,
 \left\{23q\right\}^{+-}_{1~4} \right. \nl
&+&\!\!\! 2\,(q\cdot p_6)\,q^2\,\left\{235\right\}^{+-}_{1~4} 
- 4\,(q\cdot p_2)\,(q\cdot p_3)\,\left\{65q\right\}^{+-}_{1~4} \nl 
&+&\!\!\! \left. 4\,(q\cdot p_2)\,(q\cdot p_5)\,\left\{63q\right\}^{+-}_{1~4} 
- 2\,(q\cdot p_2)\,q^2\,\left\{635\right\}^{+-}_{1~4}  
+ q^2\,\left\{6235q\right\}^{+-}_{1~4} \,\right].
\label{eqex2b}
\eqa
Therefore, by reconstructing denominators from the dot products, 
one easily see that the most complicated objects appearing in the 
scattering amplitude 
are rank one 6-point functions and rank two 5-point functions, 
as expected.\\

Some comments are in order. What I have done in the previous
examples is equivalent to Fierz reordering the spinorial strings. 
This already gives the correct structure to apply the reduction algorithm, 
without using \eqn{eqext}.

\noindent The fact that some diagrams are simpler to compute 
than others, depends on what one gets after Fierz reordering.
In fact, after elimination of all sum indexes, one can end up with momenta 
containing $q$ (that I denote with capital letters) either 
close to each other or "sandwiched" between spinors. Namely
\bqa
\left\{ABC \cdots \right\}^{+-}_{i~j} \nonumber
\eqa
or
\bqa
\left\{A\right\}^{+-}_{i~j} \, 
\left\{B\right\}^{+-}_{k~l} \, 
\left\{C\right\}^{+-}_{m~n} \cdots ~. \nonumber
\eqa
In the first case, the final result is simpler because the equation
$\rlap{/} q \rlap{/} q = q^2$ can be immediately applied, while, 
in the second case, one first needs to use \eqs{eq9} and
some $\gamma$-algebra.

\noindent In example 1 there are no singular structures, while, in example 2, 
a single quantity, that can vanish in the collinear limit, 
appears in the denominator (see eqs. (\ref{eqex2a}) and (\ref{eqex2b})). 
This roughly corresponds to a situation with a single Gram determinant in the 
denominator, that is known to be numerically quite stable \cite{glover}
(actually, since $<i j><i j>^\ast = 2\,(p_i \cdot p_j)$ 
eqs. (\ref{eqex2a}) and (\ref{eqex2b}) are even more stable). 
In addition, even in this second case, one can in principle 
get rid of all singularities by changing the reduction when the collinear
limit is approached. For example, a reduction equivalent to \eqn{eqex2b},
but with a different singular structure, can be obtained by replacing 
\bqa
&& 1\to 3~~~~2\to 4~~~3\to 5 \nl
&& 4\to 6~~~~5\to 1~~~6\to 2 \,.
\eqa
In conclusion, one has a better control on the singularities, 
with respect to traditional reduction methods, in which one would have
immediately to start with rank three 6-point functions and a higher
number of Gram determinants. 
\section{The $n$-dimensional case}
In this section, I show how the method can be applied in $n$ dimensions.

\noindent The basic idea is simple. Starting from the traces
in \eqn{eqgen} one breaks each $n$-dimensional vector
into four-dimensional and $\epsilon$-dimensional part 
\cite{bern,taekoon,veltman} ($\epsilon= n-4$).
This allows to factorize any $\epsilon$ dependence out of
four-dimensional objects, in which all properties of the
spinor calculus can still be used in order to work out the reduction, 
exactly as shown in the previous section. This procedure is
equivalent to assume a four-dimensional-helicity regularization scheme, in
which only the loop momentum is continued in $4+\epsilon$ dimensions,
while helicity states are kept four-dimensional \cite{bern1}.

\noindent Before illustrating the procedure with explicit examples,
I introduce notations and relevant formulae.\\ 

\noindent I underline quantities that live in the $n$-dimensional
space, while a tilde denotes $\epsilon$-dimensional
objects. A generic $n$-dimensional vector $\underline{v}$ can then 
be written as 
\bqa
\underline{v} = v+ \tilde v~.
\eqa
It is always possible to chose orthogonal spaces for $4$ and
$\epsilon$-vectors, such that \mbox{$v_i \cdot \tilde{v}_j= 0$}.
Besides, $\epsilon$-dimensional $\gamma$ matrices freely anticommute
with four-dimensional ones
\bqa
        \{\gamma_\mu,\tilde{\gamma}_\nu\} \,\,= 0\,.
\eqa
As for $\gamma_5$, a mathematical correct prescription would be
keeping its definition also in $n-4$ dimensions, that implies
\bqa
         [\gamma_5,\tilde{\gamma}_\mu] \,\,= 0\,.
\label{eqgamma50}
\eqa
While this causes no trouble for diagrams with only infrared or collinear 
divergences, extra finite terms, that violate the Ward identities, 
appear when ultraviolet divergences are present. At the one-loop
level, one may either adopt \eqn{eqgamma50} 
and restore the Ward identities, by
introducing appropriate finite counter terms in the Lagrangian, 
or impose, also for $\epsilon$-dimensional matrices 
\bqa
           \{\gamma_5, \tilde{\gamma}_\mu\}\,\,= \,\,0\,,
\label{eqgamma5}
\eqa
plus extra care in manipulating traces \cite{korner2}.
In the following, I always assume \eqn{eqgamma5}. To preserve
consistency, it is enough anticommuting $\gamma_5$ to the end
of a trace {\em before} performing any $\gamma$-algebra \cite{mertig}. \\

\noindent As a first example, I compute the simple vertex
diagram in fig. 4.
\bfig(300,130)
\SetScale{1}
\sof(-10,20)
\flin{130,5}{85,50}\flin{85,50}{130,95}
\wlin{20,50}{85,50}
\glin{120,15}{120,85}{9}
\LongArrow(113,43)(113,57)
\Text(109,50)[r]{$\underline q$}
\Text(15,50)[r]{$\mu$}
\Text(131,50)[r]{$\gamma$}
\Text(55,54)[b]{$W$}
\Text(134,95)[bl]{$1$}
\Text(134,5)[tl]{$2$}
\Text(100,30)[tr]{\footnotesize{\underline B}}
\Text(100,70)[br]{\footnotesize{\underline A}}
\Text(195,90)[l]{$\underline{p_A}= \underline q+p_1$}
\Text(195,75)[l]{$\underline{p_B}= \underline q-p_2$}
\Text(195,50)[l]{$D_1= \underline{q}^2$}
\Text(195,30)[l]{$D_2= (\underline q+p_1)^2$}
\Text(195,10)[l]{$D_3= (\underline q-p_2)^2$}
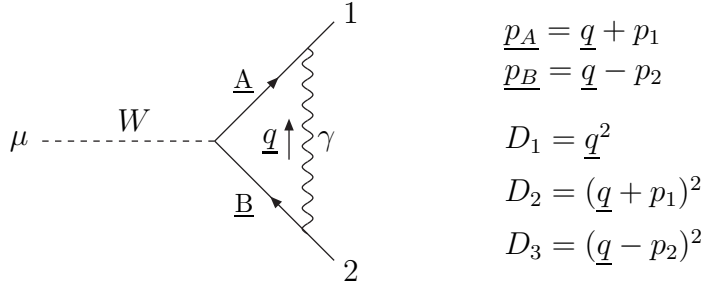
\efig{Figure 4: {\em Vertex diagram in $n$ dimensions.}}
\noindent Splitting up the $n$-dimensional traces as in ref. \cite{veltman}, 
the numerator of the amplitude can be written as follows
\bqa
{\cal A}^\mu &=&
 \left\{ \underline{\beta}\,\, \underline{A} \,\,\mu\, \Pi_-\,\, 
         \underline{B}\,\, 
         \underline{\beta} \right\}^{+-}_{1~2} = 
 \left\{ \underline{\beta}\,\, \underline{A} \,\,\mu \,\,\underline{B}\,\, 
         \underline{\beta} \right\}^{+-}_{1~2} \nl 
&=&         \left\{\beta A \mu B \beta \right\}^{+-}_{1~2} 
-\epsilon\,    \left\{A \mu B\right\}^{+-}_{1~2} 
-\tilde{q}^2\, \left\{\beta \mu \beta \right\}^{+-}_{1~2} 
+ \epsilon \tilde{q}^2\, \left\{\mu\right\}^{+-}_{1~2} \,,
\label{eqver}
\eqa
where all $\epsilon$-dimensional $\gamma$-algebra has been
explicitly worked out in order to get the desired factorization of the
$\epsilon$-dimensional objects.
Summing over $\beta$ gives
\bqa
{\cal A}^\mu &=&
         -2\,\left\{B \mu A \right\}^{+-}_{1~2} 
 -\epsilon\, \left\{q \mu q \right\}^{+-}_{1~2} 
+\tilde{q}^2\,(2+\epsilon) \left\{\mu\right\}^{+-}_{1~2}\,,
\label{eqvera}
\eqa
where the Dirac equation has also been used.

\noindent In the first term of \eqn{eqvera}, $A$ and $B$ can be extracted
with the help of \eqn{eqden}
\bqa
&&\left\{B \mu A \right\}^{+-}_{1~2} \,\,=\,\,
\frac{1}{4\,(p_1 \cdot p_2)^2} \left[\,
4\,(p_A \cdot p_2)\,(p_B \cdot p_1)\,\left\{2 \mu 1
\right\}^{+-}_{1~2} \right. \nl
&& \!\!\!\!\!\!\!\!\!\!\!\!\!\!
  \left.- 2\,(p_B \cdot p_1)\left\{2 \mu 2 q 1 \right\}^{+-}_{1~2} 
        - 2\,(p_A \cdot p_2)\left\{2 q 1 \mu 1 \right\}^{+-}_{1~2}
    +\left\{2 q 1 \mu 2 q 1\right\}^{+-}_{1~2}\,\right].
\label{eqverb}
\eqa
One easily convinces oneself that the last three terms of
\eqn{eqverb} do not contribute to the amplitude. 
In fact, performing a four-dimensional tensorial decomposition, 
each coefficient gives zero. Therefore, one gets the 
following result in terms of denominators
\bqa
\left\{B \mu A \right\}^{+-}_{1~2} \,=\,
\left\{\mu\right\}^{+-}_{1~2}\,  \left[
\frac{D_1-D_2}{p_1\cdot p_2}\,(q\cdot p_2)+2\,D_1-D_2-D_3
+2\,(p_1\cdot p_2)\right].
\eqa
Analogously 
\bqa
\left\{q \mu q \right\}^{+-}_{1~2} \,\,=\,\,
\left\{\mu\right\}^{+-}_{1~2}\,
\frac{D_1-D_2}{p_1\cdot p_2}\,(q\cdot p_2)
\eqa
and the final result reads
\bqa
{\cal A}^\mu &=& \left\{\mu\right\}^{+-}_{1~2} 
\left[ (2+\epsilon)\,\left( \tilde{q}^2+\frac{D_2-D_1}{p_1 \cdot
p_2}\,(q\cdot p_2) \right) \right. \nl
&-& \left. 4\,(p_1 \cdot p_2) + 2\,(D_3+D_2-2\,D_1) \right]\,.
\label{eqverc}
\eqa
In the previous equation, the desired reconstruction of denominators
has been achieved, but a term proportional to $\tilde{q}^2$ is also
left, that requires the evaluation of the extra scalar integral
\bqa
\tilde{I}_3 &=& \int d^nq~\frac {\tilde{q}^2}
{D_1\,D_2\,D_3} \,.
\eqa
By decomposing the integration as follows \cite{mahlon}
\bqa
\int d^nq= \int d^4q\,d^\epsilon \mu~~~~~
{\rm with}~~~~~~~~~\tilde{q}^2= -\mu^2\,
\label{decint}
\eqa
one easily gets
\bqa
\tilde{I}_3 &=& -i\frac{\pi^2}{2} + {\cal O} (\epsilon)\,,
\eqa
that is all one needs.

\noindent The result in eq. (\ref{eqverc}) can be checked by using standard
reduction techniques.\\

\noindent As a second example, I compute the box diagram in fig. 5,
contributing to the one-loop QCD corrections to 4 fermion production 
at LEP2 \cite{cc10}.
\bfig(300,130)
\SetScale{1}
\sof(-10,20)
\flin{130,5}{85,50}\flin{85,50}{130,95}
\wlin{100,66}{65,95}
\Text(61,95)[br]{$3$}
\Text(83,86)[bl]{$W^\alpha$}
\glin{20,50}{85,50}{9}
\Gluon(125,10)(125,90){-3}{9}
\Text(15,50)[r]{$\mu$}
\Text(137,50)[r]{$g$}
\Text(55,56)[b]{$Z,\gamma$}
\Text(134,95)[bl]{$1$}
\Text(134,5)[tl]{$2$}
\Text(110,80)[br]{\footnotesize{\underline A}}
\Text(90,60)[br]{\footnotesize{\underline C}}
\Text(100,30)[tr]{\footnotesize{\underline B}}
\Text(195,86)[l]{$\underline{p_A}= \underline q+p_1~~~~~
                  \underline{p_B}= \underline q-p_2$}
\Text(195,72)[l]{$\underline{p_C}= \underline q+p_1+p_3$}
\Text(195,50)[l]{$D_1= \,\,\underline{q}^2$}
\Text(195,33)[l]{$D_2= (\underline q+p_1)^2$}
\Text(195,16)[l]{$D_3= (\underline q+p_1+p_3)^2$}
\Text(195,-1)[l]{$D_4= (\underline q-p_2)^2$}
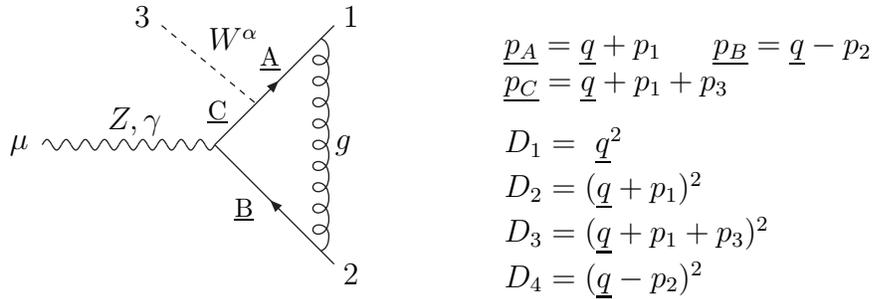
\efig{Figure 5: {\em QCD one-loop box diagram.}}
The spinorial structure reads
\bqa
{\cal A}^{\alpha \mu} &=&
 -\left\{ \underline{\beta}\,\,\underline{A} \,\,\alpha\,\,
         \underline{C}\,\,\mu\,\,\underline{B}\,\, 
         \underline{\beta} \right\}^{+-}_{1~2} \nl
&=& 2\,       \left\{B \mu C \alpha A \right\}^{+-}_{1~2}
   +\epsilon\,\left\{q \alpha C \mu q \right\}^{+-}_{1~2} \nl
  &-&2\,\tilde{q}^2\,\left[ 
          \left\{\mu \alpha A \right\}^{+-}_{1~2}
         +\left\{B \mu \alpha \right\}^{+-}_{1~2}
         +\left\{\mu C \alpha \right\}^{+-}_{1~2}
                             \right]\nl
  &-&\epsilon\,\tilde{q}^2\,\left[ 
          \left\{\alpha \mu q \right\}^{+-}_{1~2}
         +\left\{q \alpha \mu \right\}^{+-}_{1~2}
         +\left\{\alpha C \mu \right\}^{+-}_{1~2}
                             \right]\,.
\label{2ndim}
\eqa
Decomposing the integration as in \eqn{decint}, one gets
\bqa
\int d^nq~\frac{\tilde{q}^2}{D_1\,D_2\,D_3\,D_4}= 
\int d^nq~\frac{\tilde{q}^2\,q^\beta}{D_1\,D_2\,D_3\,D_4}= 
0+{\cal O}(\epsilon)\,.
\eqa
In addition 
\bqa
\int d^nq~
\frac{\left\{q \alpha C \mu q \right\}^{+-}_{1~2}}{D_1\,D_2\,D_3\,D_4}
\label{infrafin}
\eqa
does not contain any pole. In fact, infrared
divergences are absent by power counting and, in the kinematical
regions where collinear divergences take place -
namely when $q \propto p_1$ or $q \propto p_2$ - the numerator
of \eqn{infrafin} vanishes thanks to the Dirac equation. 
Therefore, in the limit $\epsilon \to 0$, only the first term 
in \eqn{2ndim} survives 
\bqa
{\cal A}^{\alpha \mu} &=& 2\,\left\{B \mu C \alpha A \right\}^{+-}_{1~2}\,.
\label{2ndima}
\eqa
Notice that, since the diagram is free from ultraviolet divergences, 
this result coincides with what one would get 
by assuming \eqn{eqgamma50}, as expected.

\noindent Terms containing up to three powers of $q$ appear in 
\eqn{2ndima}. One could in principle extract the $q$ dependence
by means of \eqn{eqden}, but this would cause the appearance of 
\mbox{$(p_1 \cdot p_2)$} into the denominator. This can be avoided as follows.
Starting, for example, from 
\bqa
 2\,\left\{q \mu q \alpha 1 \right\}^{+-}_{1~2}
\eqa
one gets, after two anticommutations 
\bqa
   2\,\left\{q \mu q \alpha 1 \right\}^{+-}_{1~2}  &=& 
                    4\,\left\{\alpha  \right\}^{+-}_{1~2}
\left[\, q^2 p_1^\mu -2\,(q \cdot p_1)\,q^\mu\, \right]
+4\,p_1^\alpha\,       \left\{q \mu q \right\}^{+-}_{1~2}
\eqa
and, rewriting the last term as a trace 
\bqa
\left\{q \mu q \right\}^{+-}_{1~2} &=& \frac{1}{N}\, {\rm Tr}[q231q\mu \Pi_+] 
\,\,=\,\, \frac{1}{N} \left[ 
 2\,(q \cdot p_2) \left\{q \mu 3 \right\}^{+-}_{1~1} \right. \nl
&-& 2\,\left. (q \cdot p_3)\,\left\{q \mu 2 \right\}^{+-}_{1~1}
          +2\,(q \cdot p_1)\,\left\{\mu q 3 \right\}^{+-}_{2~2}
    - \,q^2                \,\left\{3 2 \mu \right\}^{+-}_{1~1}
              \right] \nl
N &=& \left\{3 \right\}^{+-}_{2~1}\,,
\label{eqpar}
\eqa
from which the reconstruction of denominators can be performed as usual.

\noindent In \eqn{eqpar}, $N$ appears into the denominator, 
instead of $(p_1 \cdot p_2)$. $N$ {\em does not vanish} 
in the collinear limit $p_1 \propto p_2$, but gives zero 
when $p_3 \propto p_1$ or $p_3 \propto p_2$. 
However, since $p_3$ refers to a massive particle, those configurations
practically never occur in numerical programs, so that the described
reduction gives numerically stable expressions.

\noindent All terms in \eqn{2ndima} can be reduced as shown.
The complete result is presented in appendix A. 
Notice that the reduction can be worked out exactly in the
same way even when one of the internal fermion lines of the diagram
is massive. Only denominators change. For example, 
with the replacement $(p_1+p_3)^2 \to (p_1+p_3)^2-m^2_t$,  
the result in appendix A can also be used to compute the box diagram
that one encounters in evaluating QCD corrections 
to off-shell single top production at hadron colliders \cite{single}.\\

\noindent As a last example, in order to introduce the algorithm
in the presence of external massless vector particles, 
I show the reduction of the QCD four-point function in fig. 6, 
appearing in the calculation of the $W$ decay into 3 jets.  
\bfig(300,130)
\SetScale{1}
\sof(-10,20)
\flin{130,5}{85,50}\flin{85,50}{130,95}
\wlin{20,50}{85,50}
\Gluon(120,15)(120,85){-2.5}{9}
\Gluon(122,50)(147,50){2.5}{3}
\Text(151,50)[l]{$3,\,\alpha$}
\LongArrow(128,60)(128,74)
\Text(133,65)[l]{$\underline q$}
\Text(15,50)[r]{$\mu$}
\Text(55,54)[b]{$W$}
\Text(134,95)[bl]{$1$}
\Text(134,5)[tl]{$2$}
\Text(195,85)[l]{$D_1= \underline{q}^2$}
\Text(195,65)[l]{$D_2= (\underline q+p_1)^2$}
\Text(195,45)[l]{$D_3= (\underline q-p_2-p_3)^2$}
\Text(195,25)[l]{$D_4= (\underline q-p_3)^2$}
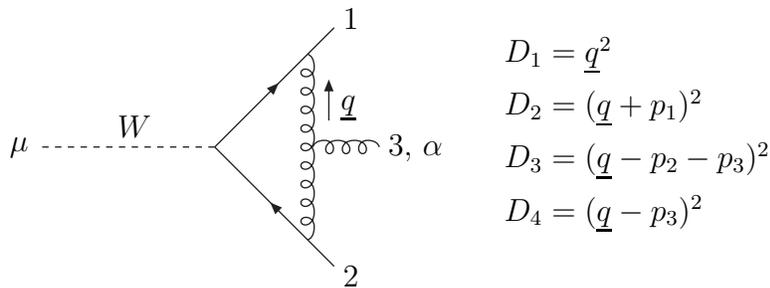
\efig{Figure 6: {\em Box diagram in $W \to 3$ jets.}}
As before, all terms in $\epsilon$ dimensions do not contribute, 
when the limit $\epsilon \to 0$ is taken, so that
only four-dimensional objects appear in the numerator of the diagram
\bqa
{\cal A}^\mu(\lambda)&=& \left\{\beta\,(q+p_1)\,\mu\,(q-p_2-p_3)\,\delta
\right\}^{+-}_{1~2}\,V^{\beta \alpha
\delta}\,\epsilon_\alpha(\lambda,p_3) \nl
V^{\beta \alpha \delta} &=& 
 g^{\beta  \alpha}(-q-p_3)^\delta
+g^{\alpha \delta}(2\,p_3-q)^\beta
+g^{\delta \beta }(2\,q-p_3)^\alpha\,.
\label{sdf}
\eqa
The polarization vector of the gluon can be expressed in terms of
spinors by means of \eqn{eqvec}. For example, by choosing
$b= p_1$ when $\lambda= -1$, one gets 
\bqa
\epsilon_\alpha(-1,p_3) &=& \frac{1}{N}\, 
\left\{ \alpha\right\}^{+-}_{3~1} \nl
N &=&\sqrt{2}\,<13>^\ast\,.
\eqa
Therefore, \eqn{sdf} can be rewritten in terms of single traces
\bqa
{\cal A}^\mu(-1) &=& \frac{2}{N} \left[\,
       \left\{p_1\,q\,\mu\,(q-p_2-p_3)\,(q+p_3)\right\}^{+-}_{3~2} \right. \nl
   &+& \left\{(q-p_2)\,\mu\,(q+p_1)\,(q-2\,p_3)\,p_1\right\}^{+-}_{3~2} \nl
&-& \left.
2\,\left\{q\,p_1\,(q-p_2-p_3)\,\mu\,(q+p_1)\right\}^{+-}_{3~2} 
    \,\right]
\eqa
and the reconstruction of denominators performed in the usual
way. For completeness, I list the result in appendix A.
\section{Conclusions}
I have presented a method to simplify the tensorial decomposition
of loop diagrams in $n$ dimensions.
Rank $l$ tensor integrals, with at least two massless momenta, can be
reduced to at most rank one tensor functions with the initial number of
denominators, plus integrals with less denominators and rank $<l$.

\noindent A better control of the numerical instabilities due 
to the appearance of Gram determinants is possible, 
with respect to traditional reduction techniques.

\noindent For processes where all heavy particles decay, the class of the 
most complicated loop diagrams lies in the domain of validity of the 
proposed method, allowing strong simplifications.

\noindent To illustrate the approach, I computed diagrams contributing to 
several processes of physical interest. 
\section*{Acknowledgment}
I wish to thank G. Passarino for stimulating and useful
discussions and for reading the manuscript.
\appendix
\section*{Appendix A}
In this appendix I present the complete reduction
of the diagrams in fig. 5 and 6.\\

\noindent The amplitude of the diagram in fig. 5 is
\bqa
~~~~~~~~~~~~~~~~
{\cal M}^{\alpha \mu} &=& \int d^nq \frac{{\cal A}^{\alpha \mu}}
{D_1\,D_2\,D_3\,D_4}\,, 
~~~~~~~~~~~~~~~~~~~~~~~~~~~~~~~~~~(A.1) \nonumber 
\nonumber 
\eqa
with ${\cal A}^{\alpha \mu}$ given in \eqn{2ndima}.

\noindent One gets, in terms of denominators and 
$N= \left\{3\right\}^{+-}_{2~1}$, the following expression
\bqa
&&\!\!\!\!\!{\cal A}^{\alpha \mu}\,\,=\,\, 
  2\,\left[
   \left\{q \mu (1+3) \alpha 1   \right\}^{+-}_{1~2} 
  -\left\{2 \mu (q+1+3) \alpha 1 \right\}^{+-}_{1~2} 
  -\left\{2 \mu (1+3) \alpha q   \right\}^{+-}_{1~2}\right] \nl &&
         +\frac{2}{N}\,\left\{
         (D_2-D_1)\,\left[
  \left\{\mu q 3 \alpha 1 \right\}^{+-}_{2~2}
 -\left\{q \mu 3 1 \alpha \right\}^{+-}_{2~2}
 -\left\{\alpha q 3 2 \mu \right\}^{+-}_{2~2}
         \right. \right. \nl &&
 +\left\{\alpha (q+1+3) \mu q 3 \right\}^{+-}_{2~2}
 +\frac{(p_1+p_3)^2}{N} \left(
  \left\{\alpha 2 3 q \mu \right\}^{+-}_{2~1}
 +\left\{\alpha 2 3 \mu q \right\}^{+-}_{2~1} 
         \right.\nl &&
         \left. \left.    
 -\left\{\mu q 3 2 \alpha \right\}^{+-}_{2~1} 
                        \right) \right]
 +(D_1-D_4)\,\left[
  \left\{q \mu 3 1 \alpha \right\}^{+-}_{1~1}
 +\left\{\alpha q \mu 2 3 \right\}^{+-}_{1~1}
 -\left\{q \alpha 2 \mu 3 \right\}^{+-}_{1~1} \right. \nl &&
             \left.
 +\left\{q \mu (q+1+3) \alpha 3 \right\}^{+-}_{1~1}
         -\frac{(p_1+p_3)^2}{N}\,
  \left\{\alpha 1 q \mu 3 \right\}^{+-}_{2~1} \right] \nl &&
 +D_1\,\left[
  \left\{\mu (1-2) \alpha 2 3 \right\}^{+-}_{1~1}
 -\left\{3 2 \mu (1-2) \alpha \right\}^{+-}_{1~1}
 -\left\{3 2 \alpha (q+1+3) \mu \right\}^{+-}_{1~1} \right. \nl &&
        \left. -\frac{(p_1+p_3)^2}{N}\, \left(
  \left\{\alpha 2 3 1 \mu \right\}^{+-}_{2~1}
 -\left\{\alpha 1 3 2 \mu \right\}^{+-}_{2~1}
         \right) \right] 
 +(D_3-D_2)\,\left[
  \left\{q \alpha 2 \mu 2 \right\}^{+-}_{1~1}
           \right.\nl &&
           \left.
 -\left\{q \mu 2 1 \alpha \right\}^{+-}_{1~1}
 -\left\{q \mu (q+1+3) \alpha 2 \right\}^{+-}_{1~1}
         +\frac{(p_1+p_3)^2}{N}\,
  \left\{\alpha 1 q \mu 2 \right\}^{+-}_{2~1} \right] \nl &&
+(p_1+p_3)^2\, \left[
  \left\{q \mu 2 1 \alpha \right\}^{+-}_{1~1}
 -\left\{q \alpha 2 \mu 2 \right\}^{+-}_{1~1}
 +\left\{q \mu (1+3) \alpha 2 \right\}^{+-}_{1~1} \right. \nl &&
      \left. \left.  -\frac{(p_1+p_3)^2}{N}\,
  \left\{\alpha 1 q \mu 2 \right\}^{+-}_{2~1} \right] \right\}\,.
~~~~~~~~~~~~~~~~~~~~~~~~~~~~~~~~~~~~~~~~~~~~~~~~~(A.2) \nonumber
\eqa
The above equation has been numerically checked against the 
result in ref. \cite{cc10}, that was obtained
using the standard reduction techniques of ref. \cite{pasve}.\\

\noindent As for the diagram in fig. 6 (with gluon polarization
$\lambda= -1$), one has
\bqa
~~~~~~~~~~~~~~~~
{\cal M}^{\mu}(-1) &=& \int d^nq \frac{{\cal A}^\mu(-1)}
{D_1\,D_2\,D_3\,D_4}\,, 
~~~~~~~~~~~~~~~~~~~~~~~~~~~~~\,(A.3) \nonumber 
\eqa
with
\bqa
&&\!\!\!{\cal A}^\mu(-1)\,\,=\,\, 
  \frac{2}{N}\,\left\{ 
      (2\,D_4-D_3)\,
 \left\{1 q \mu \right\}^{+-}_{3~2}
     +2\,D_4\,
 \left\{1 \mu q \right\}^{+-}_{3~2}
     \right. \nl &&
  +(D_2-4\,(p_1\cdot p_3))\,
 \left\{(q-2) \mu 1 \right\}^{+-}_{3~2} \nl &&
+2\,(D_1-D_2)\,\left[
 \left\{q \mu (q+1) \right\}^{+-}_{3~2}
+\left\{2 q \mu \right\}^{+-}_{3~2}
               \right] \nl &&
+2\,D_1\,\left[
 \left\{1 \mu 1 \right\}^{+-}_{3~2}
-\left\{1 2 \mu \right\}^{+-}_{3~2}
         \right]
+2\,
 \left\{2 \mu q 3 1\right\}^{+-}_{3~2} \nl &&
\left.
+2\,
 \left\{q 1 (2+3) \mu 1\right\}^{+-}_{3~2} \right\}\,, 
~~~~~~~~N=\sqrt{2}\,<13>^\ast\,.\,~~~~~~~~~~~~~~~~~~~~~~(A.4)
\nonumber
\eqa

\end{document}